\documentclass{article}
\usepackage[]{graphicx}
\usepackage[]{color}
\makeatletter
\def\maxwidth{ %
  \ifdim\Gin@nat@width>\linewidth
    \linewidth
  \else
    \Gin@nat@width
  \fi
}
\makeatother

\definecolor{fgcolor}{rgb}{0.345, 0.345, 0.345}

\usepackage{framed}
\makeatletter
\newenvironment{kframe}{%
 \def\at@end@of@kframe{}%
 \ifinner\ifhmode%
  \def\at@end@of@kframe{\end{minipage}}%
  \begin{minipage}{\columnwidth}%
 \fi\fi%
 \def\FrameCommand##1{\hskip\@totalleftmargin \hskip-\fboxsep
 \colorbox{shadecolor}{##1}\hskip-\fboxsep
     \hskip-\linewidth \hskip-\@totalleftmargin \hskip\columnwidth}%
 \MakeFramed {\advance\hsize-\width
   \@totalleftmargin\z@ \linewidth\hsize
   \@setminipage}}%
 {\par\unskip\endMakeFramed%
 \at@end@of@kframe}
\makeatother

\definecolor{shadecolor}{rgb}{.97, .97, .97}
\definecolor{messagecolor}{rgb}{0, 0, 0}
\definecolor{warningcolor}{rgb}{1, 0, 1}
\definecolor{errorcolor}{rgb}{1, 0, 0}
\newenvironment{knitrout}{}{} % an empty environment to be redefined in TeX
      \usepackage{hyperref}
\usepackage[noadjust]{cite}
\usepackage{alltt}
\usepackage[utf8]{inputenc}
\usepackage{booktabs}
\usepackage[table,x11names,dvipsnames,table]{xcolor}
\definecolor{gray}{rgb}{0.4,0.4,0.4}
\definecolor{darkblue}{rgb}{0.0,0.0,0.6}
\definecolor{cyan}{rgb}{0.0,0.6,0.6}

\makeatletter
\def\maxwidth{ %
  \ifdim\Gin@nat@width>\linewidth
    \linewidth
  \else
    \Gin@nat@width
  \fi
}
\makeatother

\definecolor{fgcolor}{rgb}{0.345, 0.345, 0.345}

\makeatletter
 {\par\unskip\endMakeFramed%
 \at@end@of@kframe}
\makeatother

\definecolor{shadecolor}{rgb}{.97, .97, .97}
\definecolor{messagecolor}{rgb}{0, 0, 0}
\definecolor{warningcolor}{rgb}{1, 0, 1}
\definecolor{errorcolor}{rgb}{1, 0, 0}

\renewenvironment{knitrout}{}{} % an empty environment to be redefined in TeX

\IfFileExists{upquote.sty}{\usepackage{upquote}}{}
\begin{document}

\title{Methods to Evaluate Lifecycle Models for Research Data Management}
\author{Tobias Weber, Dieter Kranzlmüller}
\maketitle

Lifecycle models for research data are often abstract and simple.
This comes at the danger of oversimplifying the complex concepts of research data management.
The analysis of 90 different lifecycle models lead to two approaches to assess the quality of these models.
While terminological issues make direct comparisons of models hard, an empirical evaluation seems possible. 

Lebenszyklus-Modelle für Forschungsdaten sind oft abstrakt und einfach.
Hierin liegt die Gefahr, ein zu einfaches Bild der komplexen Forschungsdaten-Landschaft zu zeichnen.
Die Analyse von 90 dieser Modelle führt zu zwei Ansätzen, die Qualität dieser Modelle zu bewerten.
Die Uneinheitlichkeit in der Terminologie erschwert einen direkten Vergleich zwischen den Modellen,
wohingegen eine empirische Evaluierung der Modelle in Reichweite liegt.

\section{Introduction}\label{sec:introduction}
Advances in science are usually the product of a team rather than individuals. 
It is obvious that more than one researcher is needed to further science,
since new insights are based on the work of others,
and scientific publications are reviewed by peers. 
Maybe less obvious is the necessity for a number of other actors:
research software developers help to develop state-of-the art tools, 
communication specialists disseminate important scientific findings,
and data librarians support researchers in data management tasks.
These three professions gain the more importance,
as the role of digital methods and forms of communication increases.

Both aspects of modern research, its collaborative nature and the fast-evolving technical possibilities, 
are best exemplified by the task to manage research data.
A large number of services, tools, protocols, best practices, and policies have been created and are currently competing for adoption. 
This state of creolization\footnote{cf. \cite{methods001}} leads itself to a research question:
How can we describe, explain, assess, and maybe even predict phenomena in research data management?  
Of what nature is the interaction between researchers and other professionals?
The most prominent answer to this question is to model research data phenomena along a lifecycle.

While the term "research data lifecycle" is used in many books, papers, blogs, a commonly shared definition is not available.
Most of these models break down the phenomena of research data management into a series of tasks or states of data and relate them to different roles or actors.
As \cite{aboutx01} indicates,
these models are often not evaluated in a manner that allows to reproducibly derive the same model for a certain purpose (explaining, educating, etc.).
A model un-evaluated is, scientifically speaking, of doubtful quality. 
If it remains unclear \emph{how} the quality of these models can be assessed,
their contribution to a better theoretical understanding of research data management remains an open question. 

The contribution of this paper is the analysis of 90 data lifecycles,
in order to identify ways to evaluate these models. 
Two approaches are presented:
\begin{itemize}
    \item One approach focuses on the comparison of data lifecycle models and tries to derive common quality indicators from the literature (and data lifecycle models published in non-classical ways.
    \item The alternative approach abstracts from the usage of the models found in the literature survey, suggests a classifcation with regard to the purposes the models is developed for and derives empirical evaluation criteria from these purposes.
\end{itemize}

The rest of the paper is structured as follows:
in \autoref{sec:relatedWork} we will examine the related work.
Our methodological approach is discussed in \autoref{sec:methods}.
The results are presented in \autoref{sec:results} and discussed in \autoref{sec:discussion}.

\section{Related Work}\label{sec:relatedWork}

\cite{dlc001} is the one of the first research data lifecycle models in the sense indicated above.
Despite the early publication date, very few practical aspects have been added to the description of research data management tasks by later lifecycle models.
It is derived from a literature review and interviews with 18 leaders of contemporary "cutting edge" projects.
Unfortunately this publication is rarely used in the literature as a reference evaluate against and check the consistency of terminology.

\cite{dlc036} shows an approach very similar to ours:
Based on a survey of lifecycle models, an abstract data lifecycle model is derived and a classification scheme is developed.
In contrast to \cite{dlc036}, we do not define a lifecycle model but a common scheme shared by all found lifecycle models.
One of features by which \cite{dlc036} classifies, is the distinction between prescriptive and descriptive models,
which comes very close to our proposal to classify along the purpose the model was designed for. 
Our method is more focused on evaluation and the resulting classification is therefore more fine-grained with regard to that. 
\cite{dlc036} provides more classifications of features, of which some are irrelevant for evaluation (e.g. the distinction betwen homogeneous and heterogeneous lifecycles).

\cite{dlc044}, \cite{dlc055} and \cite{dlc068} are alike to \cite{dlc036} in the approach to review existing models and deriving an own lifecycle model based on a gap analysis.
None of the three publications offer generic and empirical evaluation criteria or a metamodel for the existing models. 
Their lifecycle model is designed to supersede the existing approaches for a specific context. 

The model of \cite{dlc044} is not targeted at scientific data, but at open data in governmental context.
The authors clearly state the empirical methods, how the model was derived,
but the paper does not include an evaluation of the lifecycle model.

\cite{dlc055} and \cite{dlc068} both propose a lifecycle model for Big Data. 
Although they model the same phenomena, the models are not similar.
While \cite{dlc068} does not describe evaluation criteria of the model, 
\cite{dlc055} proposes the 6Vs of Big Data (Value, Volume, Variety, Velocity, Variability, Veracity) as a base to evaluate data lifecycle models in the context of Big Data.
This evaluation\footnote{carried out in \cite{dlc055eval1}}
is also applied to evaluate other data lifecycle models for their aptness to describe Big Data challenges.
This evaluation  is the most rigorous we found in the literature,
but it is limited to the context of Big Data and itself is based on a theoretical concept instead of empirical evaluation.

\cite{aboutx01} provides a scoped review of 301 articles and 10 companion documents discussing research data management practices in academic institutions between 1995 and 2016.
The review is not limited to, but includes publications discussing data lifecycle models.
The discussion includes the observation, that of the papers reviewed, only a view provided empirical evidence for their results, which is in accordance to our findings. 
The study classifies the papers based on the UK data lifecycle\footnote{\url{https://www.ukdataservice.ac.uk/manage-data/lifecycle}},
which fortunately is preserved as an attachement to this paper (its "official" version has changed since the original publication).

\section{Methods}\label{sec:methods}
A survey was executed, to derive a framework to compare data lifecycle models to each other and to find the purposes for which those models are designed for, 
Since not every research data lifecycle is described in an accademic publication\footnote{e.g. the UK data research lifecycle, see above},
our approach was to use a combination of methods of a classical literature review with a "snowball" method (following references from a first set of models to enlarge the number of found models).
Starting from a research in May 2017, which facilitated search engines (
google scholar\footnote{\url{https://scholar.google.de}},
BASE\footnote{\url{https://www.base-search.net}})
and literature databases (
ACM digital library\footnote{\url{https://dl.acm.org}}
and IEEE Xplore\footnote{\url{https://ieeexplore.ieee.org}}),
and a list of already known articles,
a first set of 35 data lifecycles was collected.

The search terms used included any combination of two out of the three following words: "research", "data" and "lifecycle".
This deliberately included lifecycle models which are not specifically dedicated to research data (e.g. governmental data, linked open data),
but we found no essential differences in both conceptualization and evaluation of these models in comparsion to research data lifecycle models in the strict sense.
Decisive inclusion criteria for a resource was a check for a textual or graphical representation of a set of actions regarding data or states of data.
Following the references to other resources (either links or citations), we stopped to collect
further models when we reached 90 data lifecycle and our analyses did not reveal new aspects.

After an evaluation of 35 lifecycle models a common pattern emerged,
which was successfully applied to the following 55 models, and therefore positively evaluated.
All models included at least one of the following characteristics, which are the building blocks of the metamodel:
\begin{itemize}
    \item A set of \textbf{states} in which data are during their scientific processing (such as creation, analysis, preservation, etc.) 
    \item A \textbf{connection} between these states (in the sense of edges in a directed graph) 
    \item A set of \textbf{roles} in the context of  to research data management (researchers, data stewards/librarians, funders, etc.)
    \item A set of \textbf{actions} with regard to research data managment (collecting, documenting, annotating, etc.) 
    \item A \textbf{mapping} of roles, actions and states to each other (e.g. "in state creation researchers describe their methods")
\end{itemize}
Since the lifecycle models differ widely with regard to their representation (differnt graphical and textual representation) a homogeneous processing was not possible at first.
To ease the analysis and comparisons between the models, they were transcribed into an XML representation.\footnote{
    These XML representations are published together with bibliographic data of the 90 lifecycle models and other resources: 10.25927/002}
A schema to validate the XML representations was used to guarantee quality of the representations. 

During the processing of the sources for the data lifecycle models, excerpts stating the purpose the lifecycle models were collected.
The classification of purposes is a result of an abstraction from these excerpts and the context in which some of the  data lifecycle models were published at (e.g. training material, service advertisements).

\subsection{Threats to Validity}
The collection method does not guarantee completeness,
which means there might be research data lifecycle models not captured by our analysis.
Since \cite{aboutx01} already provides a scoping review of the relevant literature this is an acceptable defect.
Our approach was focused on finding criteria to compare data lifecycle models to each other and to  evaluate the fitness of lifecycle models in general for certain purposes,
which does not necessitate completeness.

The list of purposes a model can be designed for is probably not complete, too, at least in a generic sense (models could for example also be used to exemplify).
But the list should include all relevant applications of models in the context of research data management.

We only included English and German resources describing data lifecycles.
As far as models had been described in other languages, they often seemed to be translations.
When German models would bias our results, we excluded them from the statistics (this is clearly stated in the text).

\begin{knitrout}
\definecolor{shadecolor}{rgb}{0.969, 0.969, 0.969}\color{fgcolor}\begin{figure}
\includegraphics[width=\maxwidth]{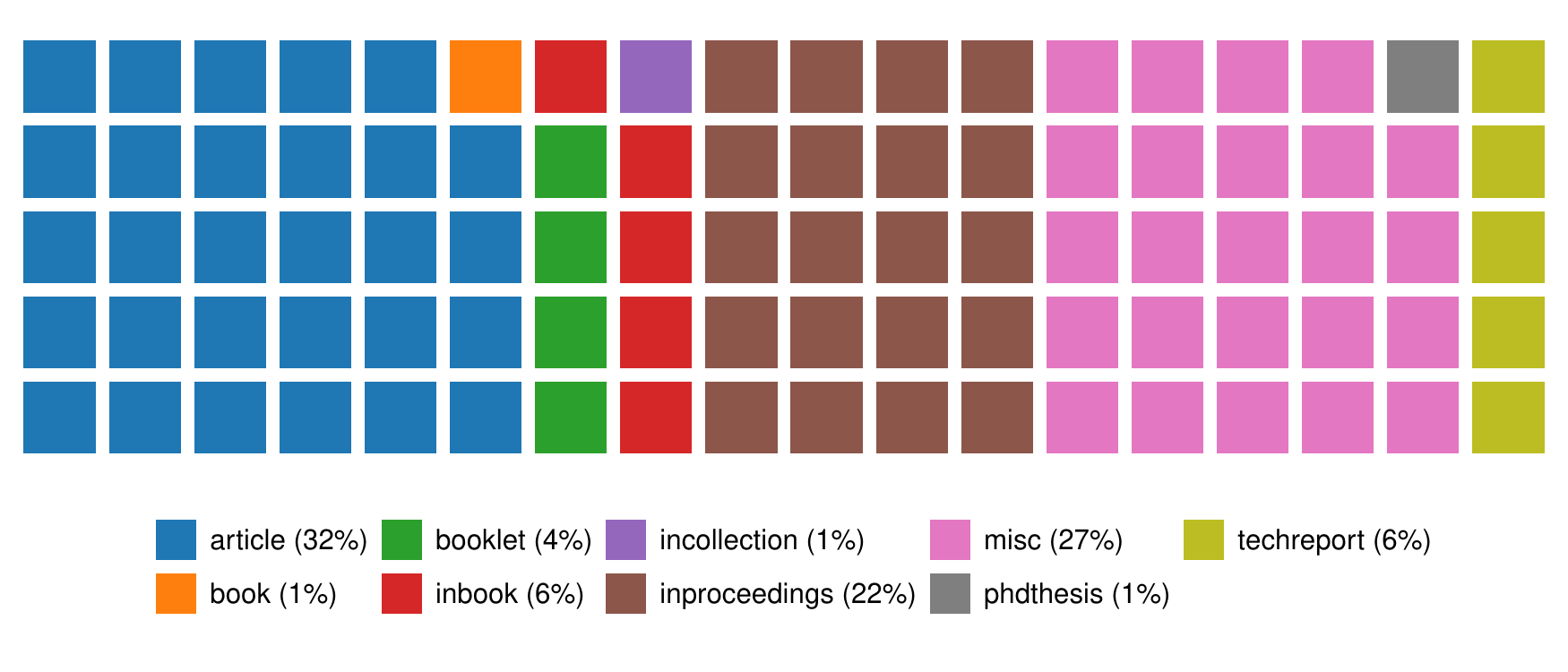} \caption[Publication types of found research data management models/lifecycles]{Publication types of found research data management models/lifecycles}\label{fig:plotPubType}
\end{figure}

\end{knitrout}
\section{Results}\label{sec:results}

The heterogenity of the sources for data lifecycle models can be seen in \autoref{fig:plotPubType}.
62\% of the models are published in a medium that is citeable in the classical sense (journals, proceedings, or books).
78\% (70) of the found models have a graphical representation.

The remainder of this section is divided into two parts:
The first part presents our statistical evaluation results, based on the metamodel presented in \autoref{sec:methods}.
The presented numbers will be the basis for the discussion how the "dimensions" of the metamodel could facilitate a comparison between data lifecycle models.
The second part proposes a classification of data life cycle models via their application, a derived evaluation method, and an example for this application. 

\subsection{Comparison of Lifecycle Models along the Metamodel}

39\% (35) of the models define actions, 14\% (13) define roles and only 13\% (12) define both.
Some of the models that only include states, "encode" an action into the state the data is currently in (e.g. "Analysing" or "data cleansing"), which makes them hard to compare with other models that separate state and actions.
11\% (10) of the models provide a mapping of activities and roles to specific states.
For partial mappings, \autoref{tab:mappings} can be consulted.

The five characteristics listed in \autoref{sec:methods} allow us to define classes of data life cycle models.
Each class extension is defined by the characteristics fulfilled by its members, i.e. there is a class for all models which define states and actions, but no roles etc.
This classification determines a partial order that allows to realise a partial comparison. 
The following data lifecycle models provide all five characteristics and are therefore members of the "highest" class with regard to the partial order:

\cite{dlc024},
\cite{dlc023},
\cite{dlc083},
\cite{dlc036},
\cite{dlc044},
\cite{dlc068},
\cite{dlc030},
\cite{dlc085}.

The number of states in the data lifecycles ranges between three and thirteen,
the number of actions between zero and 42,
the number of roles between zero and eight.
We found 399 different terms for a state, 54 different terms for roles, and 454 different terms for a research data related activity (case-insensitive matching, non-English resources were ignored).
All these numbers give evidence to the obvious heterogenity in the existing terminology in research data management.

To derive a total order from the partial order would allow us to compare all data lifecycle models to each other (and not only the classes). 
To achieve this we would need to have the criteria of completeness for each characterists,
i.e. to answer the question, whether a model includes all essential states, actions, roles, mappings and connections in the finest resolution.
Given the already stated heterogenity this task is virtually impossible to accomplish on the collected model descriptions alone:
The semantical mapping between to terms is often not possible, since they lack a rigorous definition and the models differ in their granularity.

\begin{knitrout}
\definecolor{shadecolor}{rgb}{0.969, 0.969, 0.969}\color{fgcolor}\rowcolors{2}{gray!6}{white}
\begin{table}

\caption{\label{tab:mappings}Mappings of states, actions and roles to each other (right upper half: absolute number of mappings, left lower half: percentage of mappings)}
\centering
\begin{tabular}[t]{cccc}
\hiderowcolors
\toprule
Mapping & States & Actions & Roles\\
\midrule
\showrowcolors
States & - & 35 & 11\\
Actions & 39 \% & - & 10\\
Roles & 12 \% & 11 \% & -\\
\bottomrule
\end{tabular}
\end{table}
\rowcolors{2}{white}{white}

\end{knitrout}

\subsection{Evaluation Criteria based on Model Application}
These are the classes abstracted from the 90 data lifecycle models.
Each class corresponds to the purpose a model was designed for:

\begin{itemize}
\item \textbf{Documentation:} Models can be used to describe certain aspects of reality, hence document it. 
If a model is used to document the state of research data management practices,
its main evaluation criteria is its correspondence with actual research data management practices.  
Since these practices differ widely with regard to tools, standards, protocols and policies,
there is certainly not one model that can claim to be \emph{the} research data lifecycle.
Methodologically speaking, the evaluation of a model designed to document is executed by the same approach by which it is (or should have been) created:
interviewing experts is an appropriate method to test the accurateness of such a model.
Examples for models which are used to document the actual state of research data management include the DataOne data lifecycle model\footnote{\cite{dlc004}}
and the lifecycle of CENS data.\footnote{\cite{dlc043}}

\item \textbf{Explanation:} A model explains a set of phenomena, if its usage leads to a better understanding of it.
Explanatory models are to documenting models as tutorials are to manuals.
Data lifecycle models which explain certain aspects of research data management,
are evaluated along the success in educative outcome.
The evaluation how apt a model is to explain to researchers, for example, how they can make data more reuseable,
is therefore a task that should use the methods of empirical educational theory.
The "lifecycle stages of environmental datasets" is an example of this kind of purpose.\footnote{\cite{dlc030}}

\item \textbf{Design:} Designing a desired state with a model is the (re-)arrangement of components that could be also part of a documenting model.
In the context of research data management, a model that arranges states of data items, roles, and actions can be evaluated
according to the set of features such a desired state would have.
This is comparable to the model that depicts the layout of a house: 
one can show, how this specific layout would facilitate the usage by a family, a bachelor, or old persons in need of care. 
This indicates that an evaluation of a model is only possible, if the model is assessed together with a set of objectives (a use case or a set of generic principles). 
An example for a data lifecycle model that can be subsumed under this category 
is the data lifecycle of the Inter-University Consortium for Policitical and Social Research (ICPSR).\footnote{\cite{dlc014}}

\item \textbf{Assessment:} To assess means to map the actual state to a desired state and qualify or quantify the conformance.
Either the model is used to describe the actual state or the desired state or there are two models for each of the states.
Such an assessment is implicetly carried out when statements are made that a certain service "supports the research data lifecycle".
Whether or not a set of models are suitable for assessment depends on their specific evaluation of how well they are equipped to document or to design respectively.
An example for such a usage is the United States geological survey science data lifecycle model.\footnote{\cite{dlc028}}

\item \textbf{Instruction:} Another way to relate documenting and designing models to each other,
is to use them to steer and execute transitions from the actual state to a desired state.
Such a transition typically includes the orchestration of tasks and the allocation of resources as done in classical project management.
A prominent example is to use a lifecycle model as tool to plan a data-intensive project.
Whether or not a couple of models (one documenting, one designing) is suitable for planning and executing such a transition is not only determined by the composite evaluation of the two models,
but also by the success of the transition.
An example for a research data lifecycle model that claims to support this activity is the DCC curation lifecycle model\footnote{\cite{dlc006}}
or the community-driven open data lifecycle model.\footnote{\cite{dlc044}}
\end{itemize}
\section{Discussion}\label{sec:discussion}
This section is structured in the same way as the previous one: First, the model comparisons will be discussed and after that the empirical evaluation criteria.

Models providing all five aspects of research data management should be considered of higher quality than models which only provide them partially.
While this is a first start to compare the quality between data lifecycle models, it does not take into account whether the states, actions, roles, their connections and mappings are complete.
It is obvious from the numbers presented in \autoref{sec:results} that handling of heterogenity of the terms for states, actions and roles is a very complex task.
As stated, another problem is handling the different resolutions of the lifecycle models:
there is no obvious way to handle mereological relations between states, actions and roles of two distinct models. 
A core set of states, actions and role which typically are part of a data life cycle is therefore not deducible objectively with the methods presented in this paper.
These "canonical sets" would allow to answer the question of completeness of a lifecycle model,
define a total order on the set of data lifecycle model,
and therefore a means to compare the models with regard to quality.

An option to come to such an evaluation criterion would be to postulate canonical sets.
If this turns out to be a viable option, it is recommendable to start with the 50 states, 35 roles and 84 actions that are part of the models in the highest class according to the partial order. 
A good starting point to converge the terminology would be the ontology produced by the RDA Data Foundation and Terminology Interest Group.\footnote{\url{https://smw-rda.esc.rzg.mpg.de}}

The evaluation methods proposed in \autoref{sec:results} on the other hand are ready to be used by researchers.
It is to be expected that a positive evaluation according to one purpose might imply a conflict to another one.
Take the example of documentation and explanation:
typically, good explanations place greater emphasis on certain aspects compared to others, if this helps grasping central concepts.
This might entail simplifcations or incompleteness in the model that are not acceptable in the context of documentation.

Examples for evaluating a design model with regard to objectives are Findability, Accessibility, Interoperability or Reusability (the FAIR prinicples).\footnote{\cite{general003}}
Although a convergence on these principles is a goal embraced by many,
there is no common agreement with regard to all aspects of these objectives.  
Whether these principles or maturity models\footnote{\cite{dlc080}} are more apt as a means to assess research data management practices\footnote{\cite{aboutx02}} or to instruct actors in certain data-related tasks is a question that only rigorous empirical evaluation can answer..

\section{Conclusion}\label{sec:conclusion}
Wheras a systematic comparisons of data lifecycle models is not easy, based on the approach proposed,
the evaluation criteria for models based on the purpose they were designed for is a viable option.
Scientific papers proposing a model for research data management should clearly state the purpose of the model and consequently include an evaluation with regard to this purpose.
This would bring evidence-based methods into the field of scientific infrastructure research.
Evidence-based statements improve the quality of the research, foster reproducibility of findings and ease comparison between different theoretical approaches. 
A more rigorous definition or reusage of definitions of terms will furthermore ease comparability between different models in the future. 

These considerations do not only apply for research data models,
but could be extended to other tasks of scientific infrastructure research, including, but not limited to models for research software development or standards with regard to technical scientific infrastructures. 
The improvement of methods of this research field will have impact to all disciplines,
since they will profit from new insights gained that lead to improved services of research service providers.

\section{Acknowledgements}\label{sec:acknowledgements}
We like to thank Richard Grunzke for his feedback on first ideas for this paper.
This work was supported by the DFG (German Research Foundation) with the GeRDI project (Grant No. BO818/16-1).
  \nocite{*}
\bibliographystyle{IEEEtran}
% Generated by IEEEtran.bst, version: 1.14 (2015/08/26)

\end{document}